# Towards a Cognitive Handoff for the Future Internet:

Model-driven Methodology and Taxonomy of Scenarios

Francisco A. González-Horta, Rogerio A. Enríquez-Caldera, Juan M. Ramírez-Cortés, Jorge Martínez-Carballido
Department of Electronics, INAOE
Tonantzintla, Puebla, México
{fglez, rogerio, jmram, jmc}@inaoep.mx

Eldamira Buenfil-Alpuche
Faculty of Engineering
Polytechnic University of the Guerrero State, UPEG
Taxco, Guerrero, México
eldamira@gmail.com

*Abstract*— A cognitive handoff is a multipurpose handoff that achieves many desirable features simultaneously; e.g., seamlessness, autonomy, security, correctness, adaptability, etc. But, the development of cognitive handoffs is a challenging task that has not been properly addressed in the literature. In this paper, we discuss the difficulties of developing cognitive handoffs and propose a new model-driven methodology for their systematic development. The theoretical framework of this methodology is the holistic approach, the functional decomposition method, the model-based design paradigm, and the theory of design as scientific problem-solving. We applied the proposed methodology and obtained the following results: (i) a correspondence between handoff purposes and quantitative environment information, (ii) a novel taxonomy of handoff mobility scenarios, and (iii) an original state-based model representing the functional behavior of the handoff process.

## I. INTRODUCTION

A handoff is essential to support the mobility and quality of communications on wireless networks. Its main purpose is to preserve the user communications (continuity of services or seamlessness) while different kinds of transitions occur in the network connection. The resulting handoffs pursuing such purpose are obviously single-purpose handoffs, which we claim they are not enough to face the challenges of the future Internet [1], [2], [3], and [4].

The rationale for this claim is as follows: a seamless handoff provides service continuity, but it is worthless since it works only for the specific scenario to which was stated. Therefore, a handoff should also be adaptive to any possible scenario. Now, a seamless-adaptive handoff is useless if it demands online user interventions. Consequently, a handoff should also be autonomous. Even so, a seamless-adaptive-autonomous handoff is fruitless if new security risks appear during such handoff. Thus, a handoff should also be secure. Furthermore, seamless-adaptive-autonomous-secure handoff is still unproductive if it does not perform correctly, i.e., if it does not maximize the connection time to the best available network and minimize the handoff rate. Such rationale will lead to a multipurpose handoff: seamless-adaptive-autonomous-secure-correct and thus a valuable handoff.

The development of handoffs achieving multiple desirable features has been "delayed" by the research community itself, despite it was advised since 1997 by Tripathi [1], because many authors preferred to focus on understanding and controlling very specific handoff scenarios (reductionist approach) instead of managing complex and generic handoff scenarios (holistic approach). However, recent handoff schemes, like the ones proposed by Altaf in 2008 [2] for secure-seamless-soft handovers, Cardenas in 2008 [3] for fast-seamless handoffs, and Singhrova in 2009 [4] for seamless-adaptive handoffs, show a tendency towards cognitive handoffs.

This paper presents a model-driven methodology for developing cognitive handoffs. This methodology represents the first attempt to systematically develop cognitive handoffs using a comprehensive model-based framework. The proposed methodology is founded on a synthesis of holism, reductionism, functional decomposition, model-based design, and scientific problem-solving theory.

As a result of deploying our methodology, we present a clear correspondence among cognitive handoff purposes and handoff environment information.

Besides, in order to test the resulting cognitive handoff when applying such methodology with the parameters associated to, and for a given scenario, we develop two things: i) A taxonomy of handoff mobility scenarios which gives a classification of handoff scenarios by considering all feasible combinations of several communication dimensions involved in, and ii) An original state-based model of the handoff process represented by five-state diagram which describes a general control handoff process coordinating the stages before, during, and after the handoff.

The rest of the paper is organized as follows. Section II presents the model-driven methodology we are using for developing cognitive handoffs. This section discusses the difficulties for developing cognitive handoffs and provides an overview of theoretical framework setting the basis of our methodology. Section III shows the first results we obtained from applying the methodology. These results include: (a) the correlation between context data and desirable handoff features through the definition of handoff purposes, objectives, and goals; (b) the taxonomy of handoff scenarios derived from combining all the possible transition elements involved in handoffs; and, (c) a cognitive handoff state-based model that describes a general behavior of the control handoff process. Section IV presents a basic discussion on the applicability of preliminary results. Finally, Section V concludes the paper with a summary of contributions and future work.

## II. MODEL-DRIVEN METHODOLOGY FOR DEVELOPING COGNITIVE HANDOFFS

### A. Difficulties for Developing Cognitive Handoff

The simple idea of achieving multiple purposes simultaneously is challenging even for humans. Moreover, if the intended purposes represent opposing situations which all of them are desired, then even humans need a way to balance the different purposes in conflict; e.g., the conflict between doing the job accurately and doing it quickly. In optimization theory, multi-objective optimization states that improvements to a single purpose can be made as long as the change that made that purpose better off does not make any other purpose worse off. This is called a Pareto improvement. When no further Pareto improvements can be made, then the solution is called Pareto optimal [5].

Typically, a decision-maker chooses one optimal solution according to his preference. Therefore, the first difficulty in developing cognitive handoffs arises because there are many purposes, objectives, and goals all of them in conflict that need to be tradeoff.

A second significant difficulty emerges when numerous sources of environment information need to be considered to achieve the desired multiple purposes. Six sources of context we consider include: user, terminal, network, provider, application, and handoff process. Such sources produce context data that need to be collected, transformed, and distributed at the different handoff control entities (HCEs). The challenge is how to manage large amounts of unsorted high-dimensional data that have very complicated structures and at the same time reducing the signaling traffic overload produced by this task.

The last significant difficulty is originated by the different transition elements involved in the handoff process. These elements include radio channels, base stations, IP networks, service providers, user terminals, and all the feasible combinations. This variety of elements produces a large amount of scenarios that need to be considered for an adaptive handoff scheme.

### B. Theoretical Framework

First, we state the basis for establishing our methodology.

*1) Holism and Reductionism:* Holism and reductionism are two complementary and opposing approaches for analyzing complex systems [6]. They represent different views of the relationship between the whole and the parts. Holism states that parts cannot explain the whole, the whole states the behavior of parts; i.e., it is necessary to understand how the entire handoff system determines the behavior of its components. Conversely, reductionism states that parts can explain the whole, then the behavior of parts determine the behavior of the whole. We have seen how reductionist handoff schemes achieve its goals in specific scenarios but they quickly become special cases of more general models. Holistic models are more complex models that pretend to consider all the individual parts and to understand the purposes of the whole.

*2) Model-based Design:* The model-driven paradigm has emerged as one of the best ways to confront complex systems. As it was clearly expressed by Dr. Hoffman [7], models can capture both the structure of the system (architecture) and behavior (dynamism). Model-based systems engineering [8] helps to address complexity by raising the level of abstraction, enabling developers to view system models from many perspectives and different levels of detail while ensuring that the system is consistent. The Systems Modeling Language (SysML) [7, 8] is becoming an accepted standard for modeling in the systems engineering domain. Using SysML for modeling helps to reduce ambiguity in models. In fact, models can now show the dynamic behavior of systems, including how they transition between states and how the system behaves overall.

*3) Functional Decomposition*: refers to the process of resolving a functional relationship into its constituent parts in such a way that the original function can be reconstructed from those parts by function composition. The process of decomposition [9] is undertaken for the purpose of gaining insight into the constituent components.

*4) Design as Scientific Problem-Solving:* In his inspiring paper, Braha [10] showed the similitude between the systems design process and the solving-problem process. Therefore, we developed his foundation and proposed a methodology establishing a general procedure that starts with a problem statement and ends up with the solution deployment. This theory views the problem statement as the initial state and then, by searching through a state-space, reaches a goal state representing the solution.

### C. Design and Development Procedure

Steps involved in a form of top-down procedure are:

*1) Stating the problem:* Develop a handoff procedure that can optimally achieve multiple desirable features simultaneously. The handoff procedure should be implemented for operating in real scenarios with multiple dimensions of heterogeneity. Then, as part of the problem: a) Identify and analyze the required system functions: Study the desirable handoff features that need to be implemented and determine the purpose, objectives, and goals associated to every feature. Associate a clear and single purpose to every desirable feature. Decompose each purpose into one or more objectives by identifying the performance parameters that help to quantify the achievement of every purpose. In the same way, divide every objective into one or more specific handoff goals, using optimization values and handoff context data and b) Determine the needed handoff context information: Establish what handoff criteria, handoff metrics, performance measures, handoff policies, handoff constraints, and handoff scenarios are needed to achieve every desired purpose. Study the availability, locality, dynamicity, structure, and complexity of the variables, policies, and constraints to use.

*2) Design a subsystem structure or model-based framework:* State a cognitive handoff conceptual model, i.e. identify all external context information as well as all internal context information with the highest abstraction level. Whilst internal data constitutes self-awareness, external data constitutes context-awareness of the handoff process. Then, using functional decomposition divide up the conceptual model into a number of sub-models. Every sub-model corresponds to a particular sub-problem that functionally is part of the whole handoff problem. The structure of the system may be represented with a hierarchy of models or framework enclosing the parts of the whole system organized through functional relations. Models in this framework describe the system behavior in an accurate and unambiguous way if one uses a finite set of states and a set of transition functions, thus to ease this part: Identify the associated system states and phases. These dynamic models can be formally represented using finite automata, Petri nets, timed automata, etc. [11]. The states or phases of the handoff process should describe a general behaviour rather than specific details of particular sub-models.

*3) Execute the models:* Execution of models allows verification and validation of such models. This is the difference between just drawing pictures and making pictures "live" as it was pointed out by Hoffmann in [7]. However, verification and validation should not be confused. Model verification means to test if the model satisfies its intended purposes or specifications. Model validation tests if the model provides consistent outcomes that are accurate representations of the real world. We use three strategies for these tasks: simulation, prototyping, and analysis. Whatever the strategy we choose, model testing or model checking [12] requires the use of a formal notation; e.g., modelling languages for simulation, mathematic and logic for analysis, and programming languages or middleware for model prototype implementation. If a model cannot be properly validated or verified, then it must be redesigned within the framework.

*4) Implementation stages:* Once all the models in the framework have been individually tested, the design problem now reflects a well-structured solution. A detailed design can now be generated considering the entire framework of models. This whole system design should be implemented in a whole system prototype. The final prototype is ready to be tested in-situ; should any failure occur during testing, then a review of the conceptual model or any sub-model in the framework should be performed.

*5) Solution deployment:* The cognitive handoff solution is ready to operate on a real handoff environment. The solution system (cognitive handoff) provides a simultaneous acomplishment of the multiple purposes defined by the handoff problem. Each purpose should be associated to quantitative objective functions to measure the degree in which every handoff purpose was achieved.

III. APPLYING THE MODEL-DRIVEN METHODOLOGY

*A. Purposes, Objectives, Goals, and Context Data*

The handoff context information is extensive, heterogeneous, distributed, and dynamic. It supports the whole operation of the handoff process and the achievement of multiple desirable features. From the external and internal vision of the handoff environment, we have identified five external sources of context information (creating context-awareness) and one internal source which is the handoff process itself (creating self-awareness):

*1) User context:* This context includes the user preferences, user priorities, user profiles, and user history and it is used to respond to user needs, habits, and preferences.

*2) Terminal context:* This context domain includes the following evaluating parameters: (i) Link quality: Received Signal Strength (RSS), Signal-to-Noise Ratio (SNR), Signal-to-Noise-and-Interference Ratio (SNIR), Bit Error Rate (BER), Block Error Rate (BLER), Signal-to-Interference Ratio (SIR), Co-Channel Interference (CCI), Carrier-to-Interference Ratio (CIR), etc.; (ii) Power management: Battery Type (BT), Battery Load (BL), Energy-Consumption Rate (ECR), Transmit Power in Current (TPC), Transmit Power in Target (TPT), and Power Budget (PB); (iii) Geographic mobility: Terminal Velocity (Vel), Distance from a Base Station (Dist), Geographic Location (Loc), Moving Direction (MDir), and Geographic Coverage Area (GCA). All these evaluating parameters allow the deployment of QoS-aware handoffs, power-based handoffs, and location-aided handoffs.

*3) Application context:* It includes the QoS requirements of running applications; Lost Packets (LP), Delayed Packets (DP), Corrupted Packets (CP), Duplicated Packets (DuP), Data Transfer Rate (DTR- goodput), Packet Jitter (PJ), Out-of-Order Delivery (OOD), Application Type (AppT).

*4) Network context:* This information is necessary to select among networks (before handoff), to monitor service continuity (during handoff), and to measure network conditions (after handoff) thus they are: Network Bandwidth (NBW), Network Load (NL), Network Delay (ND), Network Jitter (NJ), Network Throughput (NT), Network Maximum Transmission Unit (NMTU).

*5) Provider context*: Information about connection fees, billing models, roaming agreements, coverage area maps, security management (AAA), types of services (data, voice, video), provider preferences, and provider priorities.

*6) Handoff performance context:* This information forms the self-aware part of our cognitive model and allowing evaluation of its performance. Call Blocking (CB), Call Dropping (CD), Handoff Blocking (HOB), Handoff Rate (HOR), Handoff Latency (HOL), Decisions Latency (DLat), Execution Latency (ExLat), Evaluation Latency (EvLat), Handoff Type (HOType), Elapsed Time Since Last Handoff (ETSLH), Interruptions Rate (IR), Interruption Latency (IL), Degradations Rate (DR), Degradations Latency (DL), Degradations Intensity (DI), Utility Function (UF), Signaling Overload (SO), Security Signaling Overload (SSO), Improvement Rate (ImpR), Application

Improvement Rate (AppImpR), User Improvement Rate (UsrImpR), Terminal Improvement Rate (TermImpR), Successful Handoff Rate (SHOR), Imperative Handoff Rate (IHOR), Opportunist Handoff Rate (OHOR), Dwell Time In the Best (DTIB), Authentication Latency (AL), Detected Attacks Rate (DAR), Online User Interventions Rate (OUIR), Tardy Handoff Rate (THOR), and Premature Handoff Rate (PHOR).

Once we have identified the context data from all the context sources and the desired handoff features that we wish to implement, then, we assign a qualitative purpose to every desired feature and, a set of quantitative objectives and goals to every handoff purpose. Tables I and II summarize such previous description.

TABLE I. DESIRED FEATURES, PURPOSES, OBJECTIVES, AND GOALS

| Desired Handoff Features | Qualitative | Quantitative | |
|---|---|---|---|
| | Purposes | Objectives | Goals |
| Seamlessness | Maintain continuity of services or preserve user communications | Reduce DR, DL, DI, IR, IL | Minimize (BER, BLER, CCI, NL, ND, NJ, LP, DP, CP, DuP, PJ, TPC, TPT, ECR, CB, CD, HOB, HOL) Maximize (RSS, SNR, SNIR, SIR, CIR, NBW, NT, NMTU, DTR, BL, ETSLH) |
| Autonomy | Preserve handoff operation independent of users | Reduce OUIR | Maintain (IL < app.Timeout) |
| Security | Maintain a constant level of security along the handoff | Reduce SSO, DAR | Minimize (AL, SO, HOL) Maintain (High Encryption) |
| Correctness | Keep user always connected to the best network with minimal handoffs | Reduce HOR Increase DTIB | Minimize (HOR) Maximize (DTIB) |
| Adaptability | Keep success of all handoff objectives across any scenario | Multi-objective optimal balance Increase SHOR | Keep every desirable feature within its success range. Maximize (SHOR) |

TABLE II. OTHER DESIRED PROPERTIES OF COGNITIVE HANDOFFS

| Desired Handoff Features | Qualitative | Quantitative | |
|---|---|---|---|
| | Purposes | Objectives | Goals |
| Necessary | Prevent unnecessary handoffs | Start HO only if it is imperative or opportunist Maint. HOR = IHOR + OHOR | Imperative if (UFcurr<Thinf) Opportunist if (UFcurr>Thsup) UFtarget is SuffB & ConB |
| Selective | Avoid selecting the wrong target | Verify target is consistently better (ConB) and sufficiently better (SuffB) | SuffB: UFtarget > (UFcurr + Δ) ConB: SuffB is maintained for SP time |
| Efficient | Operate quickly and well-organized to decide how to perform the handoff (HO) | Select the best method, protocol, or strategy according to the HOType, AppType, and Mobility state. Reduce DLat, ExLat, EvLat | Define HO policies or conditions for choosing MIP, SIP, MAHO, NAHO, or other protocols |
| Beneficial | Augment benefits to applications, users, and terminals after handoff | Have a better UF after HO or a maximum improvement rate (UFnew/UFold) | ImpR >> 1 Maximize (AppImpR, UsrImpR, TermImpR) |
| Timely | Initiate a HO not tardy and not prematurely | Reduce THOR and PHOR | Maintain (DLat within its tolerance range) |

These tables represent a relevant preliminary result of the applicability of cognitive handoff methodology. On one hand, they help to reduce the ambiguity and confusion on the usability of similar handoff features because every desirable handoff feature is defined in qualitative terms (purpose) and quantitative terms (objectives and goals). On the other hand, they help to correlate context data with desirable features. For instance, from Table I, we observe that RSS is correlated with seamlessness, IL with autonomy, AL with security, etc. This correlation is intended to select the context data that is needed to support every handoff purpose.

*B. Taxonomy of Handoff Mobility Scenarios*

A second significant result obtained from the proposed model-driven methodology is a new taxonomy of handoff mobility scenarios derived from combining all the possible transition elements involved in handoffs; i.e., channels, cells, networks, providers, and terminals. This taxonomy depicts all different kinds of handoffs that are possible in real networks.

Nowadays, no handoff solution exists which comprehensively addresses the entire scale of heterogeneity. Multidimensional heterogeneity [13] is the reason for the large number of handoff scenarios. If we define a handoff scenario as an array ($d_1, d_2 …, d_n$) where $d_i$ is an instance of $D_i$ the *i*th dimension of heterogeneity and there are $|D_i|$ different ways to instantiate the *i*th dimension, then by the multiplication principle there will be $|D_1|\times|D_2|\times…\times|D_n|$ possible handoff scenarios. However, for the user mobility dimension, the array (location, velocity, direction) may have distinct values at any instant along the path with infinite paths crossing the network; therefore, the number of possible mobility scenarios is infinite. Despite of such infinite scenarios, it is important to make a classification of handoffs according to the elements involved during the transition.

The complexity and treatment for a handoff depend on the type of transition that is occurring. A handoff will require of services from distinct OSI model layers depending on the elements involved in the transition. For example, a handoff between channels of the same cell is a layer 1 handoff; a handoff between cells (base stations) is a layer 2 handoff, it is homogeneous if cells use the same wireless technology, otherwise is heterogeneous; a handoff between IP networks is a layer 3 handoff; a handoff from one provider to another or between user terminals will demand the services of layers 4-7. Fig. 1 depicts the hierarchical structure of a mobile Internet in a four-layer design (core, distribution, access, and mobile). We will use this figure to explain a handoff hierarchy that involves channels, cells, networks, providers, and terminals.

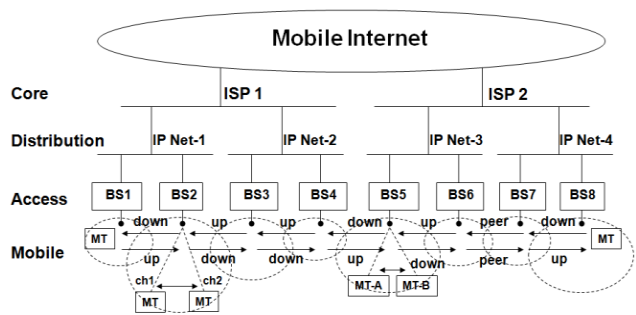

Figure 1. Hierarchy of handoff mobility scenarios. Different overlay sizes for macro, micro, pico, and femto cells.

The mobile Internet is divided into independent administrative units called Autonomous Systems (AS). An AS is a network administrated by a single organization or person. The Internet is a network of autonomous systems. Fig. 1 depicts two autonomous systems called ISP1 and ISP2 for two distinct service providers. Every ISP uses a very high-speed core network where main servers are located. Providers divide their distribution networks, physically and logically, into a number of IP networks, subnets, or VLANs (Virtual LANs), where the types of services and users are separated. Each IP Net includes a group of base stations (BS) or access points with the same or different wireless access technology. Base stations get distributed across a geographic area to offer mobile communication services. Each base station controls a cell that may have a group of channels to distribute among the associated terminals or a single channel that is shared among several associated terminals.

In Fig. 1, BS2 illustrates a layer 1 handoff when the mobile terminal (MT) changes its connection between channels ch1 and ch2 without changing of BS, IP Net, ISP, or MT. A layer 2 handoff is illustrated between BS1-BS2, BS3-BS4, BS5-BS6, and BS7-BS8. A layer 2 handoff changes from one channel to another and from one base station to another, but keeps the same IP Net, ISP, and MT; however, if the cells involved are heterogeneous, then the handoff is *vertical*, otherwise is *horizontal*. A layer 3 handoff is depicted in BS2-BS3 and BS6-BS7. A layer 3 handoff changes from one channel to another, from one cell to another, and from one IP network to another, but preserves the same provider and the same terminal; the layer 3 handoff may be heterogeneous, like in BS2-BS3, or homogeneous, like in BS6-BS7. We represent a layer 4-7 handoff, in BS4-BS5, when MT changes its communications from on channel to another, from one cell to another, from one IP Net to another, and from one ISP to another, but the user keeps the same terminal. The encryption schemes and data representation formats change from one provider to another, thus higher layer services are required. Inside the cell for BS5 we depict a handoff between terminals where the user transfers the whole session (current state of running applications) from terminal MT-A to terminal MT-B. Handoffs between terminals can be done for terminals within the same cell or different cells, within the same IP network or different IP networks, within the same provider or different providers. The terminal handoff depicted in BS5 keeps the same cell, same IP Net, and same ISP.

Fig. 2 presents a process diagram that generates the complete taxonomy of handoffs by following the different paths from the upper node to the lower nodes.

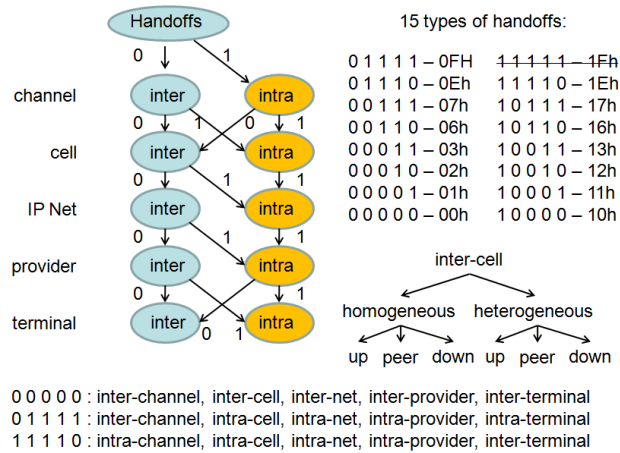

Figure 2. Generation process for handoff taxonomy. There are 15 types of feasible handoffs that can be implemented in real wireless overlay networks. The 1Fh is not a handoff.

Every handoff type in this taxonomy should be complemented or further classified according to many other criteria by using the handoff classification tree of Nasser et al. in [14].

*C. Cognitive Handoff State-Based Model*

By applying the second step of the model-driven methodology, design a subsystem structure, we created a cognitive handoff conceptual model and its first decomposition model both illustrated and discussed in [13]. Following the reductionist approach, we now focus on a major component of the handoff system, the cognitive handoff control system. At this stage, we designed a state-based model whose purpose is to understand the general behavior that should have the handoff control system. Thus, this model represents our third main result obtained from following the methodology.

Fig. 3 shows a five-state diagram modeling a general control handoff process. The states are: (1) Disconnection, (2) Initiation, (3) Preparation, (4) Execution, and (5) Evaluation. This model describes a generic control handoff system coordinating the stages before, during, and after the handoff. We describe each state briefly:

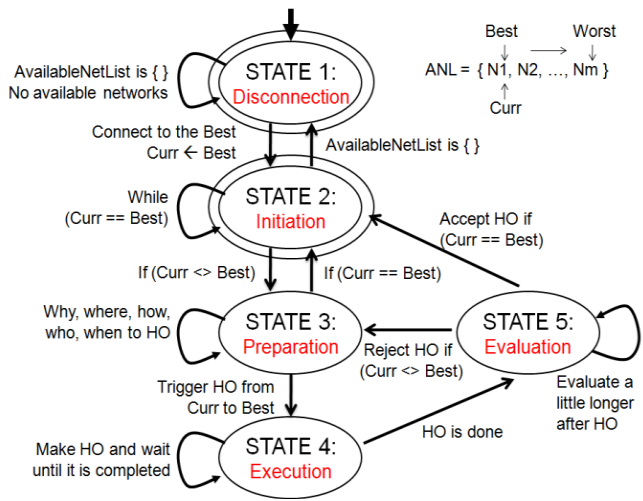

Figure 3. A handoff control model. This state diagram shows a reactive and deterministic behavior of cognitive handoffs.

*1) Disconnection:* is the initial state and one of the two final states. Here, the terminal is disconnected but discovering available networks. The process will stay here while there are no available networks.

*2) Initiation:* in this state the terminal is connected to the best available network and communications flow normally. This is another final state. The process stays here while there are no reasons (imperative or opportunistic [15]) to prepare for a handoff. If current connection breaks and no other network is available, then the process goes back to the disconnection state.

*3) Preparation*: as soon as a better network appears, the process changes to the preparation state. Here is where properly the handoff begins. This state decides why, where, how, who, and when to trigger the handoff. The handoff in progress can be rolled back to initiation if current link becomes again the best one.

*4) Execution*: once a control entity decides to trigger a handoff, there is no way to rollback; the handoff will be performed. This state knows the current and destination networks, the active application to be affected, and the strategy or method to use.

*5) Evaluation*: once the link switch is made, the control entity enters the evaluation state. This state recombines the measures for every objective function taken before and during the handoff, with new samples taken after the handoff to determine its successfulness. The evaluation latency is adjusted to a stabilization period [16].

## IV. RESULTS DISCUSSION

In this research, we have shown a new methodology to systematically develop cognitive handoffs, which are expected to be in operation in the mobility scenarios of the future Internet. Such methodology is based on a sound theoretical framework including: methods for analyzing complex systems, the model-based systems engineering, the functional decomposition approach, and the scientific problem-solving theory. There are five stages in the proposed methodology: 1) state the problem, 2) design a model-based framework, 3) execute the models, 4) implement a prototype, and 5) deploy the solution. Thus, we have presented three main results obtained from applying the first two stages of the methodology: i) a cascade relationship of desired features, purposes, objectives, goals, and context data; ii) a taxonomy of handoff mobility scenarios; and iii) a generic state-based model for a cognitive handoff control system.

Furthermore, there are some other issues that require detailed discussion: (a) the complexity of a cognitive handoff system, (b) the evaluation of cognitive handoff models, and (c) the implementation of cognitive handoffs.

### A. Cognitive Handoff Complexity

In [13] we showed two main properties of complex systems that are also present in cognitive handoffs: the hierarchic structure of systems and the property of emergence. In this section we provide other reasons of why cognitive handoffs are complex software systems: (1) Cognitive handoffs exhibit a rich set of behaviors: reactive, proactive, deterministic, non-deterministic, context-aware, self-aware, etc.; behavior is determined by the particular desirable features associated to handoffs. (2) Cognitive handoffs can be stated as multi-objective optimization problems. (3) Cognitive handoffs are driven by events in the physical world; e.g., the user mobility, the user preferences, the provider services, the coverage areas, etc. (4) Cognitive handoffs maintain the integrity of hundreds or thousands of records of information while allowing concurrent updates and queries. (5) Context information is extensive, heterogeneous, dynamic, and distributed. (6) Cognitive handoffs control real-world entities, such as the switching of data flows through a large set of available networks, providers, and terminals. (7) Handoff management has a long-life span; handoffs will exist in all future wireless networks. (8) Handoff management is a key issue for wireless industry and standardization bodies. Grady Booch in [17] provides further discussion on the attributes of complex software systems.

### B. Evaluation of Cognitive Handoff Methodology and Models

Now, as a result of applying our proposed methodology, one gets a set of models that are different in purpose (intentions), usability (applicability), notation (language), and abstraction (hierarchy).

Methodology and each model must be evaluated, either by quantitative evaluation, which comprises the definition of criteria and metrics intended to measure one specific property or, conversely by a qualitative evaluation which is related to credibility that comes from the way in which the cognitive maps are built and the clarity it represents the opinion's of most experts [18].

In relation to a qualitative evaluation of the methodology, one requires to think on the stages proposed by the development process, the kind of activities to accomplish in each stage, the strength of its theoretical basis, the kind of lifecycle in the development process, etc. Meanwhile, corresponding quantitative evaluation, metrics should be applied to all asociated parametres in the stages of the process.

With respect to evaluate models, we made a clear distinction in Section II.C between verification and validation. The verification tests if the model satisfies its purpose, whilst validation tests if the model outcomes are representations of reality. During the development process of a new system, special purpose models are built to support the understanding that goes on during the development and no hard data emerge from such models, thus, they can only be verified, but not validated.

It is worth to notice that in this paper, we deal with a specific kind of model belonging to those known as soft models [18]. Soft models are intended to understand rather than to predict and therefore verification is the way to qualitatively evaluate such models. Specifically, the theoretical framework in Section IIB has solid and proven bases.

### C. Cognitive Handoff Implementation

We envision the implementation of cognitive handoffs as a network of distributed agents cooperating and competing to take any type of handoff to success. We distinguish between agents for controlling the handoff process (HCEs) and agents for managing the handoff context data (CMAs). The CMAs are responsible for recollecting the context data and updating the handoff information base at the HCEs. CMAs are located in user terminals and distributed in different layers of the network infrastructure. HCEs are located also in every user terminal and at the network access layer; HCEs perform a handoff control process like the one depicted in Fig. 3. Thus, let us develop the state-based model as follows.

A dynamic ordered list of available networks (ANL) is organized from best to worst, according to the value of desirability calculated for every network. The desirability metric is a utility function combining a broad set of network selection criteria. The best network is the one with highest desirability. The value of desirability for the $n$th network, named $D_n(\mathbf{v})$, may have a geometric or stochastic distribution depending on the dynamic nature of context variables used as selection criteria, and arranged in a criteria vector $\mathbf{v} = (V1, V2, …, Vm)$. We use Equation (1) to represent a general mathematical model for the desirability function:

$$D_n(\mathbf{v}) = \sum(K + W_i)\log(V_i^+) - \sum(K + W_j)\log(V_j^-) \quad (1)$$

The set of decision variables $(V1, V2, …, Vm)$ fetched for the $n$th available network is partitioned in two subsets: $V_i^+$ and $V_j^-$; where $V_i^+$ is the set of criteria that contribute to the desirability (e.g. NBW and NT) and $V_j^-$ is the set of variables that contribute to the undesirability (e.g. NL and ND). $W_i$ and $W_j$ are weights corresponding to each variable such that $W_i$ and $W_j \in \Re[0,1]$, $\sum W_i = 1 = \sum W_j$ and $K$ is a scaling factor so that small changes in the context variables reflect big changes in $D_n(\mathbf{v})$.

For geometric distributions, a proactive handoff strategy may anticipate handoff decisions and for stochastic distributions a reactive handoff strategy with thresholds, hysteresis margins, and dwell-timers may prevent unnecessary handoffs. The control handoff process illustrated in Fig. 3 shows a reactive and deterministic procedure; reactive, because the process starts the preparation for a handoff until another network with higher desirability is present and, deterministic, because it is always possible to determine the current state of the process within one of five states.

Fig. 4 and Fig. 5 depict geometric distributions of desirability with different handoff strategies. Fig. 4 shows a proactive strategy where the handoff preparation starts before the target network improves the current connection. Fig. 5 shows a reactive strategy where handoff preparation starts after the target network has improved the current connection.

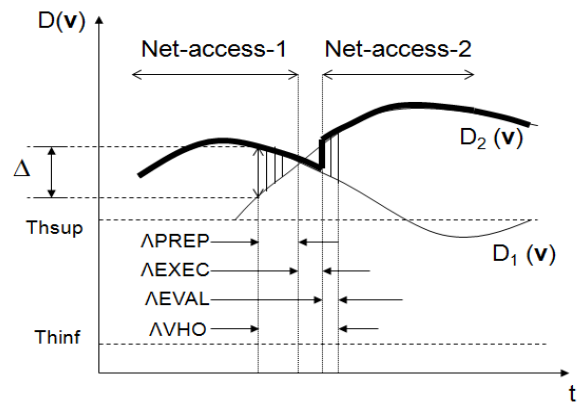

Figure 4. A proactive handoff strategy.

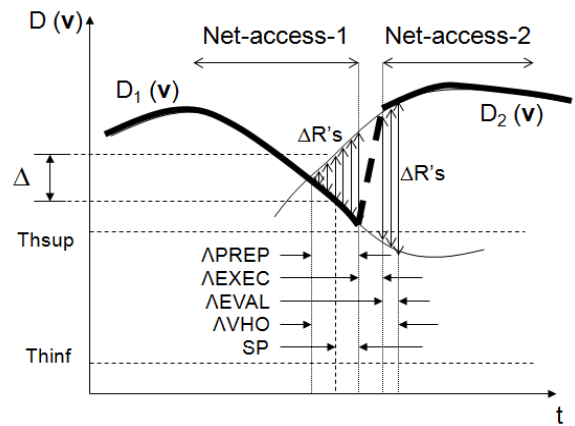

Figure 5. A reactive handoff strategy.

The darken line over the desirability functions illustrate the current connection. The performance parameters ΛPREP,

ΛEXEC, ΛEVAL, and ΛVHO depict the latencies for the different stages: preparation, execution, and evaluation. Configuration parameters include Δ (hysteresis margin), desirability threshold (Thsup, Thinf), and dwell-timer (SP). Relative Desirability measures are (ΔRs) which are equal to |Dcurr − Dbest|.

The available network list (ANL) is a data structure located at the HCEs, but continuously updated by the CMAs. When the ANL is empty, the terminal goes to the disconnection state (State 1) and stays there while such list is empty. CMAs are continuously discovering new networks and ordering the list from the highest desirable networks to the lowest desired networks.

The change from disconnection state to initiation state (State 2) occurs as soon as new networks are available. The HCE selects the best available network from the list and connects the terminal to it. The State 2 is the Always Best Connected state because the terminal will stay connected to the best network as long as no other available network improves the current connection.

The change from initiation to preparation (State 3) occurs when a new network is improving or has improved the current network. Handoff decisions, in State 3, start by identifying a reason to begin the preparation for a handoff (why). Next, selecting the target network (where). Then, deciding what strategy, method, or protocol to choose (how). Then, deciding what HCE will be responsible to trigger the handoff (who), and finally, deciding the best moment to trigger the handoff (when). The chosen handoff strategy, method, or protocol depends on the current handoff scenario (as those depicted in Fig. 1) and the type of handoff in progress (as those illustrated in Fig 2).

The decision to trigger a handoff in one terminal changes the control process from preparation to execution (State 4). The trigger handoff decision activates a procedure to change the data flows of an application from one access network to another, within specific handoff and time constraints. The switching mechanism takes a time ΛEXEC to complete.

Once the switching process is completed, the HCE enters to the evaluation state (State 5). This is an important stage of feedback to the handoff control process. At this stage, the HCE has a constrained period of time to decide to accept or reject the recently executed handoff. One condition for handoff success occurs if the new current connection is the best available connection, but others include measuring the objective functions, associated to every handoff purpose, and if all these measures are within a boundary region of acceptable quality, then the cognitive handoff is successful, otherwise it is defective and outliers should be corrected.

## V. CONCLUSION AND FUTURE WORK

Cognitive handoffs are multipurpose handoffs achieving many desirable features simultaneously; e.g., seamlessness, autonomy, security, correctness, and adaptability. The development of cognitive handoffs is a challenging task that has not been properly addressed in the literature. Therefore, we proposed a new model-driven methodology for developing cognitive handoffs. We applied the proposed methodology and obtained a clear relationship between handoff purposes and handoff context information, a new taxonomy of handoff scenarios, and an original state-based model of a generic control handoff process.

We continue developing and integrating the models generated by the cognitive handoff methodology. A future work is to organize such models in a comprehensive framework of models representing the functional issues for the whole cognitive handoff process. Further work is needed to study the availability, locality, dynamicity, structure, and complexity of variables, metrics, polices, and constraints involved in cognitive handoffs. The evaluation of the cognitive handoff methodology by quantitative techniques demands more work. We are preparing a manuscript to analyze the cognitive handoff problem as a multi-objective optimization problem using the cellular automata approach to simulate complex handoff scenarios.

### ACKNOWLEDGMENT

F.A. González-Horta is a PhD candidate at INAOE Puebla, Mexico, and thanks the financial support received from CONACYT Mexico through the doctoral scholarship 58024.